\documentclass[a4paper,11pt]{article}
\usepackage{pos}
\renewcommand{\logo}{\relax}
\title{ MEG~II experiment status and prospect}

\author*{Manuel Meucci} \author{on behalf of the MEG~II collaboration}

\affiliation{INFN Sezione di Roma,\\
Piazzale A.~Moro 2, 00185 Rome, Italy}

\affiliation{Dipartimento di Fisica dell'Universit\`a ``Sapienza'',\\
Piazzale A.~Moro 2, 00185 Rome, Italy}

\emailAdd{manuel.meucci@roma1.infn.it}

\abstract{
The MEG~II experiment at Paul Scherrer Institute (PSI) in Switzerland aims to achieve a sensitivity of $6\times10^{-14}$ on the charged lepton flavor violating decay $\mu^+\to e^+\gamma$. The current upper limit on this decay is $4.2\times10^{-13}$ at 90\% Confidence Level (CL), set by the first phase of MEG. This result was achieved using the PSI muon beam at a reduced intensity, $3\times10^7~\mu^+/$s, to keep the background at a manageable level. The upgraded detectors in MEG~II can cope with a higher intensity, thus the experiment is expected to run at a  $7\times10^7~\mu^+/$s intensity. The new low mass, single volume, high granularity tracker, together with a new highly segmented timing counter, guarantees better resolutions for the positron detection. Moreover, the replacement of the old PhotoMultiplier Tubes (PMTs) with Multi-Pixel Photon Counters (MPPCs) in the inner face of the liquid xenon photon detector improved its performance. The details of the upgraded detectors and their present status will be discussed, together with the latest results from last year's pre-engineering run and the perspective for the 2021 run, the first with all the detectors and electronics installed.
}

\FullConference{%
  *** The 22nd International Workshop on Neutrinos from Accelerators (NuFact2021) ***\\
  *** 6–11 Sep 2021 ***\\
  *** Cagliari, Italy ***}

\begin{document}
\maketitle

\section{Charged lepton flavor violation}
The Standard Model (SM) of particle physics, though being a successful theory, is not complete.
It is thought to be a low energy approximation of a more general theory yet to be discovered. The lepton flavor of charged particles is conserved in the SM, and even in its extension that includes the neutrino oscillation the rate of a process that does not conserve this quantity is $\sim10^{-54}$, non observable by a real experiment. Some Beyond SM (BSM) theories foresee the Charged Lepton Flavor Violation (CLFV) at an observable rate, thus the processes that violate this conservation law are a good probe of the SM. 

\begin{figure}[htbp!]
    \centering
    \includegraphics[width=0.99\textwidth]{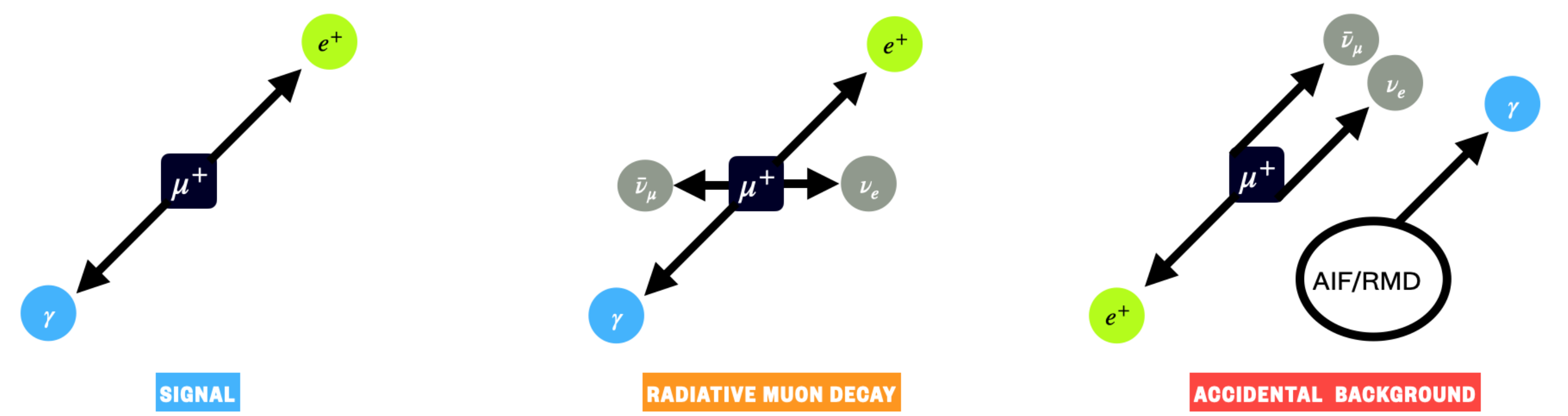}
    \caption{Schematic drawing of the relevant processes in the $\mu^+\to e^+\gamma$ search: signal, RMD background and accidental background.}
    \label{fig:signal}
\end{figure}

The $\mu^+\to e^+\gamma$ process is a CLFV process sensitive to new physics, and it is favorable because of its simple kinematics.
When the muon is at rest, in fact, the decay products are emitted back to back with the same energy $E_{e^+}=E_{\gamma}=m_\mu/2=52.8$~MeV. An experiment that wants to measure this final state has to cope with two sources of background: the Radiative Muon Decay (RMD) and the accidental background. The first one is the $\mu^+\to e^+\nu_e\bar{\nu_\mu}\gamma$ decay, that can be misinterpreted as a signal event when the two neutrinos carry away a small fraction of the energy. The second one is the accidental coincidence of a positron from the normal muon decay, the Michel decay ($\mu^+\to e^+\nu_e\bar{\nu_\mu}$), and a photon from a RMD or the Annihilation In Flight (AIF) of a positron in the experimental apparatus. Figure \ref{fig:signal} shows a drawing of the signal and background processes involved in the CLFV search in the $\mu^+\to e^+\gamma$ channel.

The best upper limit on the Branching Ratio (BR) of this decay is $4.2 \times 10^{-13}$ at 90\% CL, set in 2016 by the MEG experiment \cite{baldini2016search}.
The experiment took data from 2009 to 2013 at the PSI high intensity muon beam. Such beam can reach an intensity of $1\times10^8~\mu^+$/s, but the optimal setting for MEG was $3\times10^7~\mu^+$/s. This limitation comes from the signal and accidental background rates: the first is proportional to the beam rate, the second to its square power. Thus, increasing the beam rate is beneficial only if the detector allows for a good background rejection. 

To improve the present upper limit on the $\mu^+\to e^+\gamma$ BR the MEG collaboration planned an upgrade of the detector \cite{baldini2018design}.
Improving the overall performance of the subdetectors allows for a better rejection of the background events, and consequently makes possible the use of the muon beam at a higher intensity. The goal of the upgraded experiment, which is called MEG~II, is to collect 3 years of data at $7\times10^7~\mu^+$/s to reach a sensitivity of $6\times10^{-14}$ at 90\% CL \cite{baldini2021search}.

\section{\texorpdfstring{$\mu^+\to e^+\gamma$}{Muegamma} search with MEG~II detector}
\begin{figure}[htbp!]
    \centering
    \includegraphics[width=0.55\textwidth]{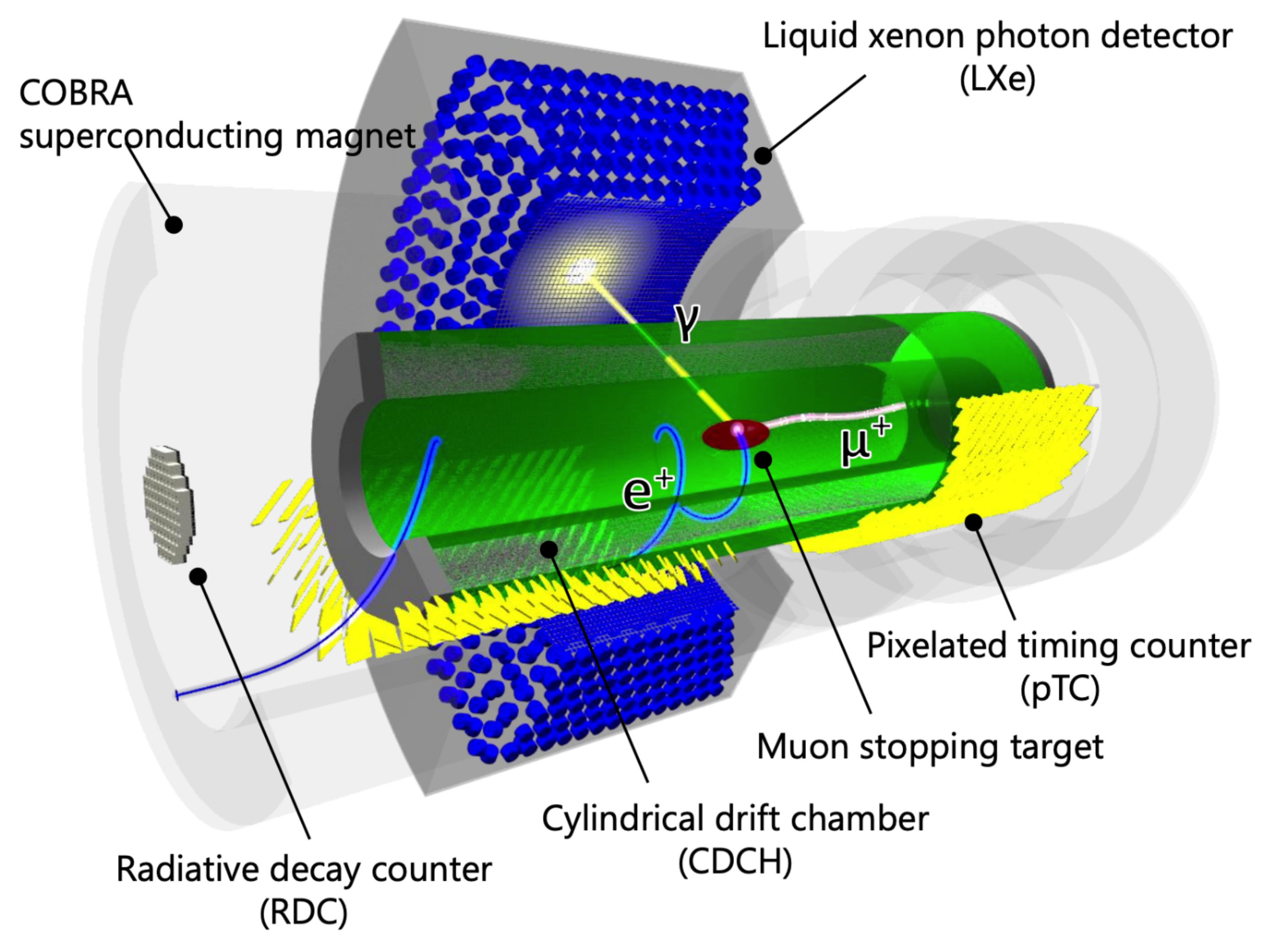}
    \caption{Drawing of the MEG~II detector}
    \label{fig:meg2}
\end{figure}

The MEG~II detector is shown in figure \ref{fig:meg2}.
Its main subdetectors are a Liquid Xenon photon detector (LXe), a magnetic spectrometer for the positron detection and a Radiative Decay Counter (RDC) as a RMD background tagging detector. The magnetic spectrometer is composed of a Cylindrical Drift CHamber (CDCH) for tracking and energy measurement, and a pixelated Timing Counter (pTC) for timing measurement. The spectrometer is placed inside a COnstant Bending RAdius (COBRA) magnet that generates an axially graded magnetic field with an intensity ranging from 0.49~T to 1.27~T.

The trigger signal for the positron has to be issued by a fast detector with excellent time resolution that can provide prompt time and direction information. Since the CDCH is affected by the latency introduced by the drift time, the pTC was designed to carry out this task \cite{de2016high}. It is composed of two identical modules, one placed DownStream (DS) and another placed UpStream (US) the muon stopping target, each module containing 256 scintillating tiles read out by Silicon PhotoMultipliers (SiPMs). The good time resolution is a consequence of the fine granularity provided by the presence of such high number of tiles, that guarantees a high number of hits from a positron track.

The positron tracker is a single volume ultra light cylindrical drift chamber \cite{baldini2020ultra}.
This detector performance is strongly affected by the multiple scattering, so being light is one of its crucial characteristics. It has 1728 20~$\mu$m thick golden tungsten sense wires and 40/50~$\mu$m thick silver plated Al cathode wires, with a cathode:anode ratio of 5:1. The CDCH is filled with a light gas mixture of 90:10 He:iC$_4$H$_{10}$ plus a small concentration of isopropyl alcohol and oxygen. The total radiation length crossed by a positron is $1.58\times10^{-3}$~X$_0$ for each turn.

The photon detector is a large tank filled with 900L of liquid xenon \cite{ogawa2017liquid}.
It measures the energy, time and conversion point of the incoming photon with high precision, using the liquid xenon as active material. The readout is managed by 630 2~inches PMTs sensitive to Vacuum UltraViolet (VUV) light at T=165~K and, on the inner face, by 4092 VUV-sensitive $12\times12$~mm$^2$ MPPCs. The latter are insensitive to the magnetic field and ensure a finer granularity of the detector, improving efficiency and spatial resolution.

\section{X(17) boson measurement}
\begin{figure}[htbp!]
    \centering
    \includegraphics[width=0.4\textwidth]{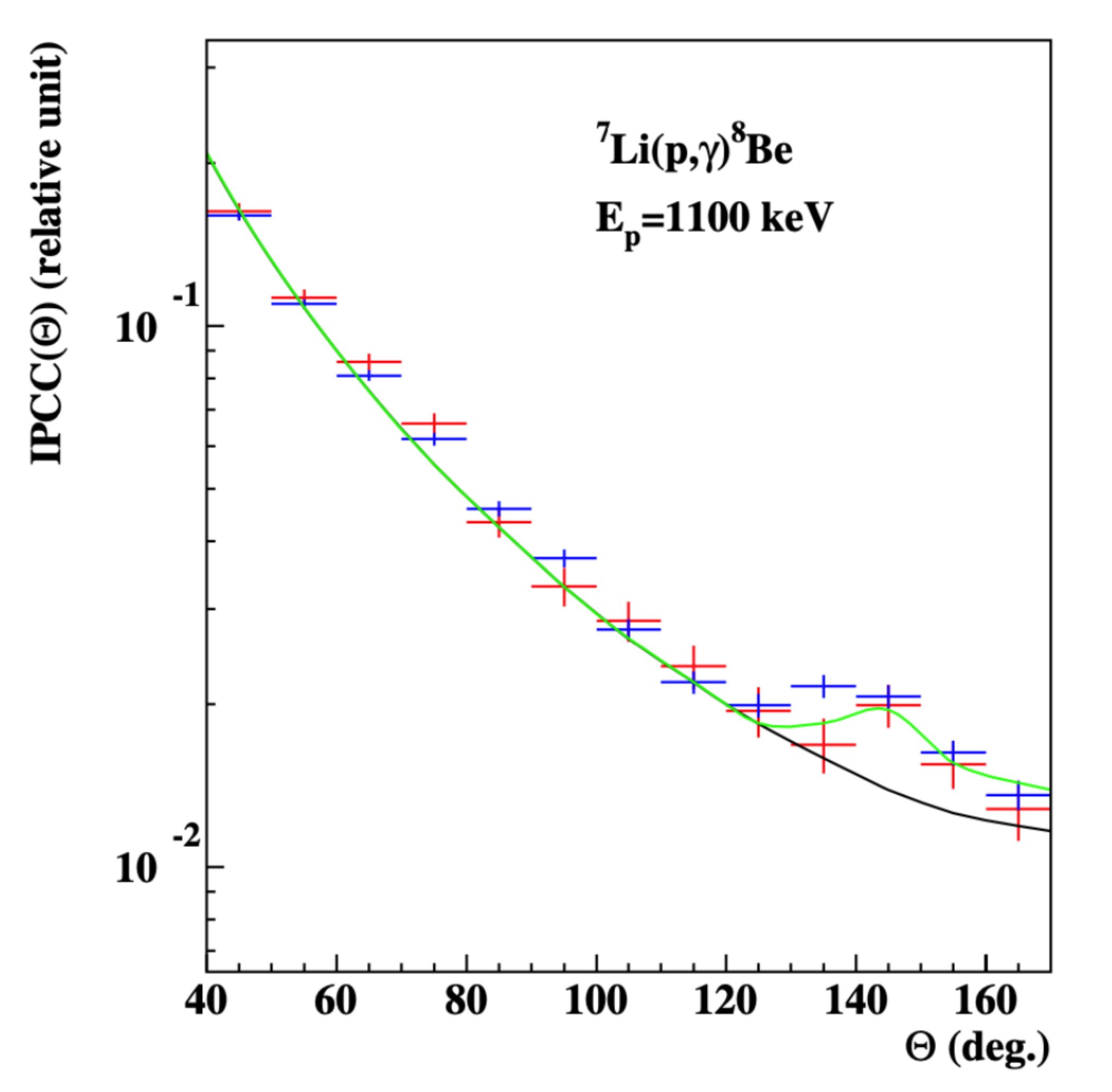}
    \caption{Angular distribution of the IPC pairs from the $^7{\rm Li}({ p},{e}^+{e}^-)^8{\rm Be}$ measurement at Atomki \cite{krasznahorkay2018new}.}
    \label{fig:xboson}
\end{figure}

In 2016 a measurement at the Atomki laboratory (Debrecen, Hungary) observed an excess in the angular distribution of the Internal Pair Creation (IPC) in the nuclear reaction $^7{\rm Li}({ p},{e}^+{e}^-)^8{\rm Be}$ \cite{krasznahorkay2018new}, as shown in figure \ref{fig:xboson}. 
This anomaly was confirmed by following measurements, even in the $^3{\rm H}({p},{e}^+{e}^-)^4{\rm He}$ reaction \cite{krasznahorkay2020new}. An interpretation of this measurement is that the excess is due to the production of a new physics boson, mediator of a fifth fundamental force that describes the interaction between dark matter and ordinary matter \cite{feng2016protophobic}. 

The mass of this hypothetical particle is 17~MeV, hence its name X(17).
Repeating the IPC measurement with an independent experiment can confirm that the observed anomaly is not an artifact of the detector geometry, as proposed in \cite{aleksejevs2021standard}, where the anomaly is explained within the SM. The MEG~II experiment is going to repeat the $^8$Be measurement in 2022 with a better invariant mass resolution and angular acceptance, using a proton beam from a CW accelerator available for LXe calibrations and the magnetic spectrometer for the ${e}^+{e}^-$ tracking. To maximize the resolution the CW target region was redesigned in order to minimize multiple scattering. This new setup is incompatible with the LXe calibration and cannot be used during the MEG~II muon beam data taking period, but a compatible prototype has been built and will take some preliminary data during the end of the 2021 muon run.

\section{2020 and 2021 runs}
The commissioning of the MEG~II subdetectors is ongoing since 2017, using 2-5 months of muon beam time each year.
During the 2020 run all the subdetectors were installed and powered, and only a small part of them was read out because of the limited availability of the readout electronics. That years' muon beam run was characterized by the study of the whole experiment stability under full intensity muon beam and by the first MEG~II run of LXe calibration with the Charge EXchange (CEX) reaction with a pion beam ($\pi^-+p\to\pi^0n,\,\pi^0\to\gamma\gamma$).

The full readout electronics was installed during the first half of 2021, followed by the complete Trigger and Data AcQuisition (TDAQ) system commissioning.
This will allow to test the subdetectors fully equipped for the first time in the 2021 run and to further study the performance of the whole detector. The final goal of this run is to take the first physics data of the experiment and to test the preliminary setup for the X(17) measurement \cite{chiappini2021towards}.

The LXe observed a decrease of the Photon Detection Efficiency after the exposure to the muon beam, most likely due to surface damage induced by the radiation.
This issue was studied and confirmed in the 2020 run, and a possible solution has been identified in the thermal annealing of the MPPCs. This procedure can be carried out by heating the photo sensor with hot water or with Joule heat during the beam shutdown periods programmed in the first months of every year. In the same run the first CEX calibration took place, and it was possible to estimate the energy and absolute time resolution of the subdetector at $E_\gamma=55~$MeV using the small amount of read out channels available. The goal of the 2021 run is to repeat the CEX calibration with the full electronics and to further study the PDE degradation to optimize the solution to adopt for the full MEG~II data taking period.

The RDC proved to be able to efficiently tag RMD events during 2020 run and is then ready to take physics data in 2021.
The working module is installed DS, but another module to be installed US is under development. This will improve RMD detection, allowing to tag low energy positrons that are emitted towards the US side of the detector. This module has to cope with a high rate environment, and a Resistive Plate Chamber (RPC) with Diamond-Like Carbon (DLC) resistive electrodes prototype has been tested under muon beam in 2020. The further test foreseen in the 2021 run will help to define the final design of this subdetector.

The pTC is ready for the physics run and successfully tested the stability under the pion beam used for the CEX during the 2020 run. 
The goal for the 2021 run is to take data with the full detector equipped.

The CDCH faced instabilities during its commissioning, and during the 2020 run several gas mixture were tested to reach a stable condition under muon beam exposure.
The instabilities were caused by high currents on the wires due to discharges in the gas volume. These currents disappeared after the subdetector conditioning with O$_2$ and isopropyl alcohol. The optimal gas mixture was found to be 90:10 He:iC$_4$H$_{10}$+0.5\% O$_2$+1\% isopropyl alcohol, which allowed to stably take data under full intensity muon beam during the last week of 2020 run. The main goal of the 2021 run is to further test the CDCH stability and to evaluate its performance with the full readout available.

\section{Conclusion}
The MEG~II experiment, upgrade of MEG, aims to look for the CLFV decay $\mu^+\to e^+\gamma$ with a sensitivity of $6\times10^{-14}$ in three years of data taking.
Its commissioning is ongoing and all the detectors are installed. The 2020 run was crucial in the identification of the main problems and the development of their solutions: the CDCH instabilities were cured by the use of additives in the gas mixture, and the PDE degradation of the LXe MPPCs can be cured with annealing. The RDC and pTC are ready for the physics run, and the readout electronics has been fully installed in the first half of 2021, followed by the commissioning of the complete TDAQ system. The experiment will then be able to take the first physics data at the end of the 2021 run, tracing the path towards the improved $\mu^+\to e^+\gamma$ measurement.

\end{document}